# Observation of Skyrmions at Room Temperature in Co$_2$FeAl Heusler Alloy Ultrathin Films

Sajid Husain[1], Naveen Sisodia[1], Avinash Kumar Chaurasiya[2], Ankit Kumar[3], Serkan Akansel[3], Anjan Barman[2], P. K. Mudulli[1], Peter Svedlindh[3], and Sujeet Chaudhary[1]*

[1]Thin Film Laboratory, Department of Physics, Indian Institute of Technology Delhi, New Delhi 110016, India

[2]Department of Condensed Matter Physics and Material Sciences

S. N. Bose National Centre for Basic Sciences

Block – JD, Sector – III, Salt Lake, Kolkata – 700106, India

[3]Department of Engineering Sciences, Uppsala University, SE-75121, Uppsala, Sweden

*sujeetc@physics.iitd.ac.in

## Abstract

Magnetic skyrmions are topological spin structures having immense potential for energy efficient spintronic devices. However, observations of skyrmions at room temperature are limited to patterned nanostructures. Here, we report the observation of stable skyrmions in unpatterned Ta/Co$_2$FeAl(CFA)/MgO thin film heterostructures at room temperature and in zero external magnetic field employing magnetic force microscopy. The skyrmions are observed in a trilayer structure comprised of heavy metal (HM)/ferromagnet (FM)/Oxide interfaces which result in strong interfacial Dzyaloshinskii-Moriya interaction (*i*-DMI) as evidenced by Brillouin light scattering measurements, in agreement with the results of micromagnetic simulations. We also emphasize on room temperature observation of multiple skyrmions which can be stabilized for suitable choices of CFA layer thickness, perpendicular magnetic anisotropy, and *i*-DMI. These results open up a new paradigm for designing room temperature spintronic devices based on skyrmions in FM continuous thin films.

A novel magnetic pattern in which the magnetic moments exhibit a characteristic swirling configuration is referred to as a skyrmion[1,2,3,4,5]. Remarkably, the skyrmionic structure is topologically stable/protected[6] such that once formed it cannot be continuously distorted into a ferromagnet (FM) or any other magnetic state(s). Due to these properties, skyrmions have immense potential for energy efficient spintronic devices, such as skyrmion based ultra-dense storage, race track memory, magnetic random access memory and skyrmion based spin torque oscillator etc[7,8,9]. In addition to skyrmions, other magnetically ordered complex spin configurations, *viz.*, vortices and magnetic bubbles have also been studied intensively at the nanoscale by the magnetic community[10,11,12]. Though skyrmions were first theoretically predicted in the early sixties, their experimental evidence had to wait until 2009 when their presence was reported in non-centrosymmetric B20 compounds like MnSi[1], FeCoSi[13], and FeGe[14]. In all these cases, the samples were cooled to a low temperature and/or required the presence of a high magnetic field to reveal the existence of the skyrmionic state. It was understood that the non-centrosymmetric structures present in these compounds resulted in a non-centrosymmetric spin exchange or the so-called Dzyaloshinskii–Moriya interaction (DMI)[15,16]. From the atomic scale perspective, the DMI energy (EDM) is expressed as[17] $E_{DM} = \sum_{i,j} \vec{D}_{ij} \cdot (\vec{S}_i \times \vec{S}_j)$ where $\vec{D}_{ij}$ is the DMI exchange constant characterizing the interaction between the nearest neighbor spins possessing the atomic moments $\vec{S}_i$ and $\vec{S}_j$. Initially, the DMI was considered to be of bulk origin. However, it was first proposed by Fert *et al*[18] that the DMI can be induced in a heavy metal doped system or later by Bogdanov *et al*[19] in multilayer structures by breaking the inversion symmetry at the interfaces between ferromagnetic and nearby nonmagnetic ultrathin layers. The so-induced DMI is referred to as interfacial Dzyaloshinskii-Moriya Interaction (*i*-DMI). Since the most striking effect induced by *i*-DMI is the spatial twist of the magnetization leading to the formation of skyrmions even at room temperature[9,20], thus, the *i*-DMI is expected to be more effective than the bulk DMI in the case of ultrathin films.

Several techniques have been employed to directly confirm the existence of skyrmions. These include sophisticated techniques such as neutron scattering, Lorentz transmission electron microscopy, X-ray holography, etc[1,21,22,23]. The existence of topological spin structures was also revealed by the topological Hall effect[24,25]. The studies clearly suggest that the topological nature of the skyrmions makes them very exciting for applications in spintronic devices, primarily due to the giant reduction in the threshold current density by five to six orders of magnitude compared to the current density needed in currently employed FMs for driving the domain walls[26,27]. Fert's group has theoretically predicted the formation of skyrmions by means of micromagnetic simulations in confined media[4]. This non-trivial magnetic spin structure so formed is usually defined through a skyrmion number (topological charge), $N_{sk} = \frac{1}{4\pi} \int n \frac{\partial n}{\partial x} \times \frac{\partial n}{\partial y} dxdy$, where *n* is the magnetization unit vector. Based on the magnetization revolution structure, the value of the topological charge $N_{sk}$ equals to 1 for skyrmions, e.g., spin spiral[1] or hedgehog type[28].

Recently, various groups have *focused* on the ultrathin magnetic film/nonmagnetic heavy metal interface[9,29] to explore the formation of skyrmions. In studies like these, there exists plenty of room for varying the various control parameters, *viz.,* material, thickness and the structure of the interfacial layer employed for the control of *i*-DMI, exchange and anisotropy constants, etc.

By appropriate tuning of the control parameters, the strength of *i*-DMI can be varied significantly[30]. Rohart *et al.* theoretically predicted the confinement of skyrmions in Pt/Co/AlO$_x$ ultrathin films nanostructures exhibiting *i*-DMI[31]. In a recent report, room temperature magnetic skyrmions in Co/Pd[21] and Pt/Co/MgO[20] ultrathin magnetic multilayer structures were reported. To the best of our knowledge, the existence of skyrmions in high spin-polarized materials, i.e., full Heusler alloys such as Co$_2$FeAl (CFA) which possess extremely small damping constant (~0.001) among the Heusler alloys family[32] is yet to be reported. Despite the rapidly increasing number of experimental studies for gaining further insight into understanding the origin of skyrmions, the robust formation of skyrmions at room temperature in continuous thin films continues to be a challenging task.

In this article, we present the direct imaging of stable skyrmions at room temperature by employing magnetic force microscopy (MFM) in trilayer stacks containing ultrathin CFA full Heusler alloy thin films. MFM has been used in the past for the observation of skyrmions, but in all studies, the images were recorded at low temperatures or in patterned structures[33]. In the present study, a Heusler alloy ultrathin film based heterostructure Ta/CFA(*t*)/MgO has been used in truly unpatterned form (see supplementary information S1 for sample growth and the measurement details). The motivation for this work lies in the prediction that *i*-DMI originates at the interface and is inversely proportional to the FM layer thickness[30,34,35]. The micromagnetic simulations performed in the present study also corroborate the existence of skyrmions in these trilayer ultrathin film structures due to the presence of *i*-DMI. The simulations clearly show that realization of the skyrmionic state is indeed conceivable as a result of the competition between the *i*-DMI, perpendicular magnetic anisotropy and the exchange interaction in Ta/CFA/MgO thin film structure.

**Ultrathin films with Large DMI at Ta/CFA/MgO Interface (Quantitative assessment of *i*-DMI)**

We employ Brillouin light scattering (BLS) to determine the *i*-DMI in Ta/CFA/MgO ultrathin films. The advantage of using the BLS technique is that the wave-vector of the spin wave is uniquely determined by the wavelength and the angle of incidence of the laser beam and it can detect propagating spin-wave excitations simultaneously at +*k* and –*k* wave-vectors (Stokes and anti-Stokes processes). The signature of *i*-DMI in BLS is manifested as an asymmetry in the spin-wave dispersion relation[36,30] for nonreciprocal propagation of Damon-Eshbach (DE) spin-waves[37] where the spin-wave wave-vector and magnetization both lie in the sample plane and are mutually perpendicular. The *i*-DMI energy density can be determined either by modeling the full spin-wave frequency vs. wave-vector dispersion with a modified dispersion relation for DE spin waves by introducing the DMI term or from the frequency difference (*Δf*) between spin-waves with opposite (+*k* and -*k*) wave-vectors, which is given by[36],

$$\Delta f(k_x) = f(k_x) - f(-k_x) = \frac{2\gamma k_x D}{\pi M_s} \quad (1)$$

where *Δf* is the frequency difference, $k_x$ is the *x*-component of the wave vector, $\gamma$ is the gyromagnetic ratio, *D* is the *i*-DMI vector and $M_s$ is the saturation magnetization. In our case, we have chosen the latter method for determining *D*, as the estimation of *D* here is primarily determined by the experimentally measured quantities *Δf*, $k_x$ and $M_s$. Figs. 1(a) and 1(d) show typical BLS spectra, recorded from Ta/CFA(*t*)/MgO thin film heterostructures with *t*=10 and 5 nm, respectively. It is clearly observed that the positions of both the Stokes and anti-Stokes peaks in the BLS spectra move to the higher frequency values with increase in $k_x$. $\Delta f$ was extracted from Lorentzian fits to the BLS spectra and is plotted vs. $k_x$ in Figs. 1(c) and 1(f). By fitting the experimental results by using equation (1) the strength of *D* is found to be (0.06 ± 0.01) mJ/m$^2$ in Ta(10)/CFA(10)/MgO(2) and 0.2 ± 0.01 mJ/m$^2$ in Ta(10)/CFA(5)/MgO(2). It can be seen that the *i*-DMI strength increases by a factor of 4 when the thickness of CFA is reduced by half. The *i*-DMI value of 0.2±0.01 mJ/m$^2$ in Ta(10)/CFA(5)/MgO(2) is significantly larger as compared to the reported values for other systems with a similar thickness of the magnetic layer[30,38,39]. We were unable to obtain reasonable BLS signal in ultrathin CFA films (< 5 nm), possibly due to the increased damping constant and/or rotation of the anisotropy in a direction perpendicular to the plane of the CFA layer. However, the strength of *i*-DMI is known to be inversely proportional to the thickness of the FM layer. Hence, we anticipate a substantially higher value of *i*-DMI in CFA full Heusler alloy ultrathin films.

**Room Temperature Magnetic Skyrmions: Magnetic Force Microscopy (MFM) Imaging**

Figure 2(a) shows the MFM image (magnetic contrast) of the Ta(10)/CFA(1.8)/MgO(2) thin film recorded at room temperature in remnant state (zero external magnetic field) (MFM measurements details are explained in the supplementary information S2). The image displays a well-developed characteristic feature, i.e., bright regions surrounding the small circular dark/black spots. Such features reveal the presence of chiral structure in the remnant state of the CFA films. These structures are similar to those reported previously by some other groups[1,9,20]. It is emphasized here that in MFM, only the *z*-component of magnetization contributes to the MFM contrast[33] Analogous magnetic configurations were previously reported for patterned nanodisk shaped structures in their remnant state[11] and also in presence of magnetic fields in other reports[40,41]. The line (lateral) scan profiles were performed for the confirmation and characterization of the size and the periodicity of the chiral nanostructures observed in the Ta/CFA/MgO thin film. Fig. 2(c) shows the zoomed area labeled as 1 in Fig. 2(a) for a better view of the chiral structure and Fig. 2(d) shows the corresponding 3D plot where the different color contrast corresponds to the different orientations of the magnetic flux. In the following, we will show that these structures are skyrmions. The line-scan profiles can be fitted by considering the 360° domain wall profile[42,20]. The MFM phase-shift $\Delta\phi(r)$ (which produces the MFM contrast) of a 360° domain wall profile depends on the polar angle $\theta(r)$ of the magnetization as:

$$\Delta\phi(r) = C \cos\left(\phi_{tip} - \theta(r)\right) \qquad (2)$$

$$\text{where } \theta(r) = 2\tan^{-1}\left[\exp\frac{r}{\Delta}\right](r - d/2) + 2\tan^{-1}\left[\exp\frac{r}{\Delta}\right](r + d/2) \ .$$

Here *C* is a constant, $\phi_{tip}$ is the polar angle of the tip magnetization, $\Delta$ is the domain width, and *d* is the diameter of the chiral structure. Fig. 2(e) shows the line scan performed on the single domain structure within the region labeled as 1 in Fig. 2(a) (circles represent data points). The obtained profile was fitted with the equation (2). From the fitting (red line) shown in Fig. 2(e), the rotational sense of the magnetic contrast is confirmed and based on this analysis, we identify them as skyrmions in unpatterned Ta/CFA/MgO thin films at room temperature and in the remnant state (zero field). Also, the typical size of the skyrmion (for selected linear scan) is found to be ~145.5±5.2 nm. Fig. 2(b) presents the skyrmion size statistics as a histogram and the average size of the magnetic skyrmions is found to be sub-100nm (line is the Gaussian fit to all data points obtained by line profiling). Even smaller sized skyrmions are clearly observed in the low FM thickness sample, i.e., Ta(10)/CFA(1.0)/MgO(2) (please see supplementary information S2). Later we will discuss, on the basis of micromagnetic simulations, the evolution of the skyrmions in the presence of DMI. It may be pointed out that the observed size of skyrmions in Ta(10)/CFA(1.8)/MgO(2) is quite large as compared to the lateral resolution (~30 nm) of the employed MFM set-up. This allows us to observe skyrmions at room temperature. Line scan profiles on multiple skyrmions in the regions 2 and 3 in Fig. 2(a) provide the magnetization periodicity and fitting with a simple sine function yields a periodicity of 105 nm. The observed rotational sense of magnetization in these skyrmions suggests the existence of non-zero topological (winding) number which is evaluated using the micromagnetic simulations. In the following section, we discuss the results of the simulations in detail.

**Micromagnetic Simulations: Skyrmion Formation at Room Temperature**

To validate the experimental interpretation of room temperature skyrmions, micromagnetic simulations were carried out using the Mumax3 simulation code[43] (see details in supplementary information S6). As pointed above, although the *i*-DMI could be measured only in slightly thicker films [i.e., 0.2±0.01 mJ/m$^2$ for CFA(5 nm)] but it is expected to increase inversely with the FM layer thickness[30,31,20]. Hence, the simulations were performed by scanning the *i*-DMI strength in a wider range, *i.e.*, up to 2.5 mJ/m$^2$. We have used the saturation magnetization $M_s$ = 838 kA/m as extracted from SQUID measurements. Thus, the parameters employed in the micromagnetic simulations are physically viable as they correspond to the experimentally investigated unpatterned Ta/CFA/MgO thin film. The simulations successfully realized the existence of stable single skyrmions of sub-100nm diameter (data not shown here) as well as of the multiple skyrmions in an extended Ta/CFA/MgO thin films (area~2×2 μm$^2$) at room temperature (see the supplementary information S6 for more details on the simulation parameters). Fig. 3 shows the simulated multiple skyrmions in zero magnetic field at 300 K temperature. Fig. 3(a) provides a clear formation of multiple skyrmions in an extended area which clearly demonstrate features in accordance with our experimental findings (Fig. 2). The zoomed areas in Figs. 3(b) and (d) are clearly depicted as Néel type skyrmion structures which are predominantly observed due to the interfacial DMI[44,20]. Fig. 3(e) shows the line scan profile on a single skyrmion (shown by dotted line in Fig. 3(b)) together with a fit using equation (2) for determining the size and the magnetization orientation. The diameter is found to be sub~100 nm and the statistical size distribution of the skyrmions is shown in Fig. 3(c), which reveals that a large number of skyrmions lies in the range of sub-100 nm. The line scan profile and the 3D surface plot (Fig. 3(e)) where each color represents a different orientation of the magnetization reveals a chiral structure of the magnetization identical to the experimental one (Fig. 2(d)). Figs. 3(e) and (f) offer the wisdom that the magnetic moments point downward in the center of the skyrmion (black region) and that a reorientation of the magnetic moments occurs as one moves towards the periphery of the skyrmion where the magnetic moments point upwards. The topological charge $N_{sk}$ is found to be 1 for these structures. It is to be noted that skyrmion clusters have been previously demonstrated both experimentally[3,45,46] as well as theoretically[47,48] in patterned structures. These theoretical studies show that the single, double and even multiple skyrmions may coexist together via competing ferromagnetic exchange and DMI both of which are short-range interactions. However, in previous experimental results, the skyrmion clusters were observed at a lower temperature and in patterned structures, while here we show the formation of skyrmion clusters at room temperature in an extended thin films.

**Numerical Interpretations**

It is widely understood that if the proper conditions prevail with respect to exchange interaction, *i*-DMI, magnetic anisotropy energy, etc., the skyrmion formation will occur. To obtain the equilibrium configuration of magnetization in the skyrmion, Rohart *et al* derived the micromagnetic energy functional to explain the transition from FM to chiral state/configuration which is mainly controlled by *i*-DMI in ultrathin films and the size of the skyrmion is governed by the *i*-DMI strength[31]. They stated that the stability of single skyrmions can be adjudged by comparing the value of *i*-DMI with its critical limit $D_C = \frac{4}{\pi}\sqrt{AK_{eff}}$ (*A* and $K_{eff}$ are the exchange and anisotropy energy constants, respectively) in ultrathin films. In our case, the value of *i*-DMI is estimated to be about 1.2 mJ/m² by extrapolating BLS results obtained from measurements on comparably thick CFA films to the ultrathin thickness range and is also well supported by the simulations. We note that since the expected value of $D_C$ = 2.37mJ/m² is larger than our experimentally determined value of *i*-DMI, leading to $\sigma(D) > 0$, the stability of a single skyrmion is questionable. However, in our case the skyrmions are distinctly observed and verified by the simulations.

For the validation, primarily we adopted the approach demonstrated by Olivier *et al*. that the magnetostatic energy and curvature energy (which accounts for scaling of the exchange energy as 1/film thickness)[31] need to be considered in the stabilization/formation of the skyrmionic state, where the film thickness must be equal to or greater than the characteristic length $\sigma(D)/\mu_0 M_S^2$ (Ref. 21). The characteristic length is found to be ~3.5 nm using our parameters which is smaller than our experimental value of the film thickness. This supports the stability of the skyrmionic state with existing (experimental and simulation) parameters. Secondly, according to Rohart *et al*, the diameter of the skyrmions can be defined as $d \approx \sqrt{A/[2K_{eff}(1-D/D_c)]}$ [31]. In this expression, if *i*-DMI (*D*) approaches $D_c$, then the diameter of the skyrmions will become infinitely large, while if *D>Dc* then the equation of *d* is no longer valid. This is consistent with the observation of skyrmions in the present case wherein *D<Dc* and is clearly also supported by the simulations yielding the existence of multiple skyrmions.

**Conclusions**

In summary, the observation of room temperature skyrmions in ultrathin Ta/Co$_2$FeAl/MgO trilayers is reported based on magnetic force microscopy imaging supported by micromagnetic simulations and Brillouin light scattering measurements. The MFM line scan profiling clearly reveals the nanoscale chiral structure and the structure periodicity consistent with the formation of skyrmions. The topological state has been confirmed by micromagnetic simulation by considering a suitable combination of interfacial Dzyaloshinskii-Moriya *(i*-DMI) and anisotropy energy magnitudes. The study highlights the prospects of room temperature generation of skyrmions in continuous thin films without requiring the additional cost and time intensive film patterning constraints. The skyrmions observed in these CoFe$_2$Al (1.8 nm) films were of the sub-100nm size and even sub-50 nm in thinner Co$_2$FeAl (1.0 nm) films. It is envisaged that this robust formation of skyrmions at room temperature is a significant experimental breakthrough and will pave the way towards the industrial development of the new field of skyrmitronics (skyrmion+electronics) for the fabrication of the skyrmion based devices for ultra-dense memory storage and logic applications etc.

**References:**

1. Mühlbauer, S. *et al.* Skyrmion lattice in a chiral magnet. *Science* **323,** 915–919 (2009).
2. Münzer, W. *et al.* Skyrmion lattice in the doped semiconductor $Fe_{1-x}Co_xSi$. *Phys. Rev. B* **81,** 041203(R) (2010).
3. Nagaosa, N. & Tokura, Y. Topological properties and dynamics of magnetic skyrmions. *Nat. Nanotechnol.* **8,** 899–911 (2013).
4. Sampaio, J., Cros, V., Rohart, S., Thiaville, A. & Fert, A. Nucleation, stability and current-induced motion of isolated magnetic skyrmions in nanostructures. *Nat. Nanotechnol.* **8,** 839–44 (2013).
5. Fert, A., Cros, V. & Sampaio, J. Skyrmions on the track. *Nat. Nanotechnol.* **8,** 152–156 (2013).
6. Kravchuk, V. P. *et al.* Topologically stable magnetization states on a spherical shell: Curvature-stabilized skyrmions. *Phys. Rev. B* **94,** 144402 (2016).
7. Hanneken, C. *et al.* Electrical detection of magnetic skyrmions by non-collinear magnetoresistance. *Nat. Nanotechnol.* **10,** 1039–1043 (2015).
8. Siracusano, G. *et al.* Magnetic radial vortex stabilization and efficient manipulation driven by the Dzyaloshinskii-Moriya interaction and spin-transfer torque. *Phys. Rev. Lett.* **117,** 87204 (2016).
9. Woo, S. *et al.* Observation of room-temperature magnetic skyrmions and their current-driven dynamics in ultrathin metallic ferromagnets. *Nat. Mater.* **15,** 501–506 (2016).
10. Im, M.-Y. *et al.* Symmetry breaking in the formation of magnetic vortex states in a permalloy nanodisk. *Nat. Commun.* **3,** 983 (2012).
11. Moutafis, C. *et al.* Magnetic bubbles in FePt nanodots with perpendicular anisotropy. *Phys. Rev. B* **76,** 104426 (2007).
12. Moutafis, C., Komineas, S., Vaz, C. A. F., Bland, J. A. C. & Eames, P. Vortices in ferromagnetic elements with perpendicular anisotropy. *Phys. Rev. B* **74,** 214406 (2006).
13. Yu, X. Z. *et al.* Real-space observation of a two-dimensional skyrmion crystal. *Nature* **465,** 901–904 (2010).
14. Yu, X. Z. *et al.* Near room-temperature formation of a skyrmion crystal in thin-films of the helimagnet FeGe. *Nat. Mater.* **10,** 106–109 (2011).
15. Dzyaloshinsky, I. A thermodynamic theory of 'weak' ferromagnetism of antiferromagnetics. *J. Phys. Chem. Solids* **4,** 241–255 (1958).
16. Moriya, T. Anisotropic superexchange interaction and weak ferromagnetism. *Phys. Rev.* **249,** 91 (1960).
17. Rohart, S., Miltat, J. & Thiaville, A. Path to collapse for an isolated Néel skyrmion. *Phys. Rev. B* **93,** 214412 (2016).
18. Fert, A. & Levy, P. M. Role of anisotropic exchange interactions in determining the properties of spin-glasses. *Phys. Rev. Lett.* **44,** 1538–1541 (1980).
19. Bogdanov, A. & Rößler, U. Chiral symmetry breaking in magnetic thin Films and

multilayers. *Phys. Rev. Lett.* **87,** 37203 (2001).

20. Boulle, O. *et al.* Room-temperature chiral magnetic skyrmions in ultrathin magnetic nanostructures. *Nat. Nanotechnol.* **11,** 449–455 (2016).

21. Pollard, S. D. *et al.* Observation of stable Néel skyrmions in Co/Pd multilayers with Lorentz transmission electron microscopy. *Nat. Commun.* **8,** 14761 (2017).

22. Büttner, F. *et al.* Dynamics and inertia of skyrmionic spin structures. *Nat. Phys.* **11,** 225–228 (2015).

23. Park, H. S. *et al.* Observation of the magnetic flux and three-dimensional structure of skyrmion lattices by electron holography. *Nat. Nanotechnol.* **9,** 337–342 (2014).

24. Neubauer, A. *et al.* Topological Hall effect in the a phase of MnSi. *Phys. Rev. Lett.* **102,** 186602 (2009).

25. Yokouchi, T. *et al.* Formation of in-plane skyrmions in epitaxial MnSi thin films as revealed by planar Hall effect. *J. Phys. Soc. Japan* **84,** 104708 (2015).

26. Yu, X. Z. *et al.* Skyrmion flow near room temperature in an ultralow current density. *Nat. Commun.* **3,** 988 (2012).

27. Everschor, K., Garst, M., Duine, R. A. & Rosch, A. Current-induced rotational torques in the skyrmion lattice phase of chiral magnets. *Phys. Rev. B* **84,** 64401 (2011).

28. Zhou, Y. *et al.* Dynamically stabilized magnetic skyrmions. *Nat. Commun.* **6,** 8193 (2015).

29. Emori, S., Bauer, U., Ahn, S.-M., Martinez, E. & Beach, G. S. D. Current-driven dynamics of chiral ferromagnetic domain walls. *Nat. Mater.* **12,** 611–616 (2013).

30. Cho, J. *et al.* Thickness dependence of the interfacial Dzyaloshinskii–Moriya interaction in inversion symmetry broken systems. *Nat. Commun.* **6,** 7635 (2015).

31. Rohart, S. & Thiaville, A. Skyrmion confinement in ultrathin film nanostructures in the presence of Dzyaloshinskii-Moriya interaction. *Phys. Rev. B* **88,** 184422 (2013).

32. Husain, S., Akansel, S., Kumar, A., Svedlindh, P. & Chaudhary, S. Growth of $Co_2FeAl$ Heusler alloy thin films on Si(100) having very small Gilbert damping by Ion beam sputtering. *Sci. Rep.* **6,** 28692 (2016).

33. Milde, P. *et al.* Unwinding of a skyrmion lattice by magnetic monopoles. *Science* **340,** 1076–1080 (2013).

34. Kim, N. H. *et al.* Improvement of the interfacial Dzyaloshinskii-Moriya interaction by introducing a Ta buffer layer. *Appl. Phys. Lett.* **107,** 142208 (2015).

35. Torrejon, J. *et al.* Interface control of the magnetic chirality in CoFeB/MgO heterostructures with heavy-metal underlayers. *Nat. Commun.* **5,** 4655 (2014).

36. Nembach, H. T., Shaw, J. M., Weiler, M., Jué, E. & Silva, T. J. Linear relation between Heisenberg exchange and interfacial Dzyaloshinskii–Moriya interaction in metal films. *Nat. Phys.* **11,** 825–829 (2015).

37. Damon, R. W. & Eshbach, J. R. Magnetostatic modes of a ferromagnet slab. *J. Phys. Chem. Solids* **19,** 308–320 (1961).

38. Gross, I. *et al.* Direct measurement of interfacial Dzyaloshinskii-Moriya interaction in X|CoFeB|MgO heterostructures with a scanning NV magnetometer (X=Ta,TaN, and W). *Phys. Rev. B* **94,** 64413 (2016).
39. Belmeguenai, M. *et al.* Brillouin light scattering investigation of the thickness dependence of Dzyaloshinskii-Moriya interaction in CoFe ultrathin films. *Phys. Rev. B* **93,** 174407 (2016).
40. Shinjo, T. Magnetic vortex core observation in circular dots of permalloy. *Science* **289,** 930–932 (2000).
41. Stebliy, M. E. *et al.* Experimental evidence of skyrmion-like configurations in bilayer nanodisks with perpendicular magnetic anisotropy. *J. Appl. Phys.* **117,** 17B529 (2015).
42. Romming, N., Kubetzka, A., Hanneken, C., Von Bergmann, K. & Wiesendanger, R. Field-dependent size and shape of single magnetic skyrmions. *Phys. Rev. Lett.* **114,** 177203 (2015).
43. Vansteenkiste, A. *et al.* The design and verification of MuMax3. *AIP Adv.* **4,** 107133 (2014).
44. Moreau-Luchaire, C. *et al.* Additive interfacial chiral interaction in multilayers for stabilization of small individual skyrmions at room temperature. *Nat. Nanotechnol.* **11,** 444 (2016).
45. Yu, X. *et al.* From the cover: Magnetic stripes and skyrmions with helicity reversals. *Proc. Natl. Acad. Sci.* **109,** 8856–8860 (2012).
46. Zhao, X. *et al.* Direct imaging of magnetic field-driven transitions of skyrmion cluster states in FeGe nanodisks. *Proc. Natl. Acad. Sci.* **113,** 4918–4923 (2016).
47. Rózsa, L. *et al.* Skyrmions with attractive interactions in an ultrathin magnetic film. *Phys. Rev. Lett.* **117,** 157205 (2016).
48. Leonov, A. O., Robler, U. K. & Mostovoy, M. Target-skyrmions and skyrmion clusters in nanowires of chiral magnets. *EPJ Web Conf* **75,** 5002 (2014).

**Acknowledgments**

Two of the authors (SH and AKC) acknowledge Department of Science and Technology, Govt. of India for providing INSPIRE Fellowship. Authors thank the HPC and NRF facilities of IIT Delhi for computational resources and MFM measurement, respectively. AB thanks, Department of Science and Technology, Govt. of India for funding under grant no. SR/NM/NS-09/2011(G). This work was in part supported by Knut and Alice Wallenberg Foundation (KAW) Grant No. KAW 2012.0031, Sweden.

Figures Caption:

Figure1| *i*-DMI in Ta/CFA/MgO ultrathin layers. (a) and (d) show the BLS spectra of Ta(10)/CFA(10)/MgO(2) and Ta(10)/CFA(5)/MgO(2) thin films, respectively, recorded for various $k_x$-values. The magnetic field ($H$ = 0.1T) was applied in the sample plane and perpendicular to the plane of incidence of the laser beam. (b) and (e) provide the $\Delta f$ (frequency the difference of the two peaks) for one of the $k_x$-value such as $1.81\times10^7$ m$^{-1}$. (c) and (f) show the $\Delta f$ vs. $k_x$ which provide the *i*-DMI strength from its slope (given in the inset). [Lines in (b) and (e) are fit to the Lorentzian function and in (c) and (f) red lines correspond to fits using equation (1)].

Figure2| Typical magnetic force microscopic (MFM) image of the Ta(10)/CFA(1.8nm)/MgO(2) thin film displaying room temperature magnetic skyrmions in the remnant state (zero external magnetic field). (a) MFM image as recorded in large scan area. (b) The statistical distribution of the size of the magnetic skyrmions obtained by manual line scan profiling (line is the Gaussian fit to all data points). (c) The zoomed view of a single skyrmion labeled as 1 in (a). (d) A 3D surface plot of an experimentally observed skyrmion (different colors reveal different orientations of the magnetic flux). Lateral line scan profiling in (e) and fitted with equation (2), (f) and (g) represent the *magnetization periodicity* corresponding to the label 2 and 3 marked in (a). (Open circles represent the experimentally obtained data points from the line scan profiling. The dark cyan line in (f) and (g) are the fits using a sine function for visualization of the periodicity (with period=105 nm) of the magnetization in the case of multiple/clusters of skyrmions).

Figure3| (a) Micromagnetically simulated stable skyrmions in an extended thin film area of 2×2μm$^2$, (b) zoomed view of multiple skyrmions and (c) the statistical distribution of the skyrmion size in a whole simulated area. (d) Ultra-zoomed (shown by dotted line in (b)) view for visualization of a single skyrmion indicating the formation of distinctly visible skyrmion. This observation of distinct spin texture qualitatively matches with the experimentally found skyrmions. (e) Line scan profile of a single skyrmion (shown by dotted line in Fig. 3(b)) and the red line is fit to the equation (2) for determining the size and the magnetization orientation (open circles are the simulated points obtained using line scan profiling). (b) The 3D surface plot of a simulated skyrmion (extracted from (b) indicated by dotted line) (different colors correspond to the different polar angles of the magnetization).

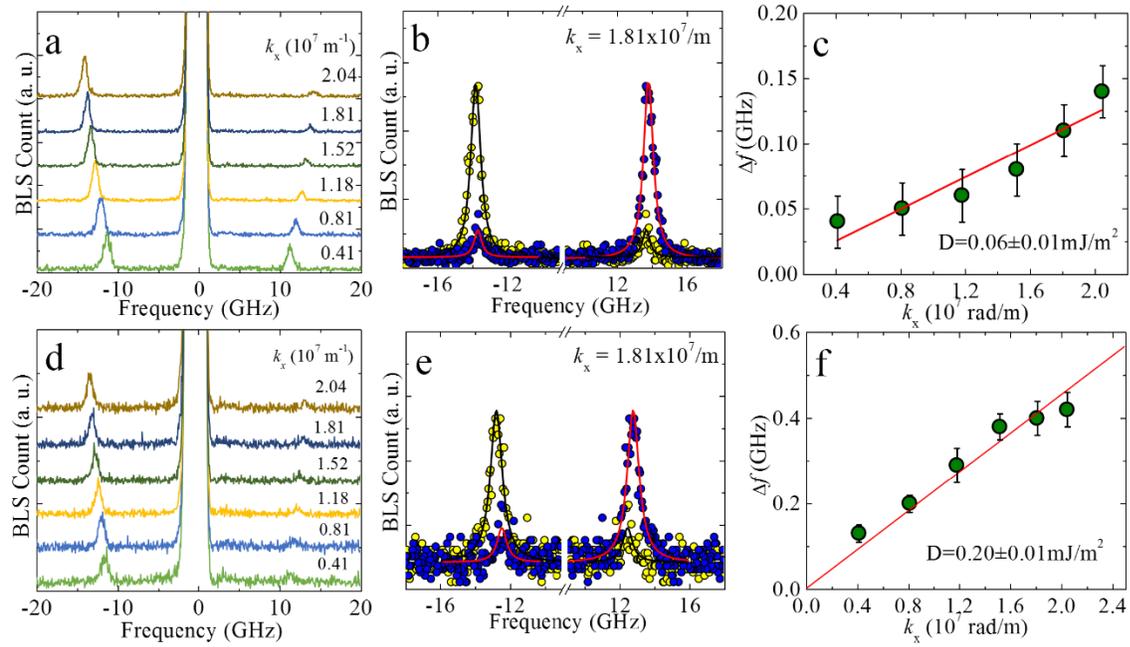

**Figure 1**

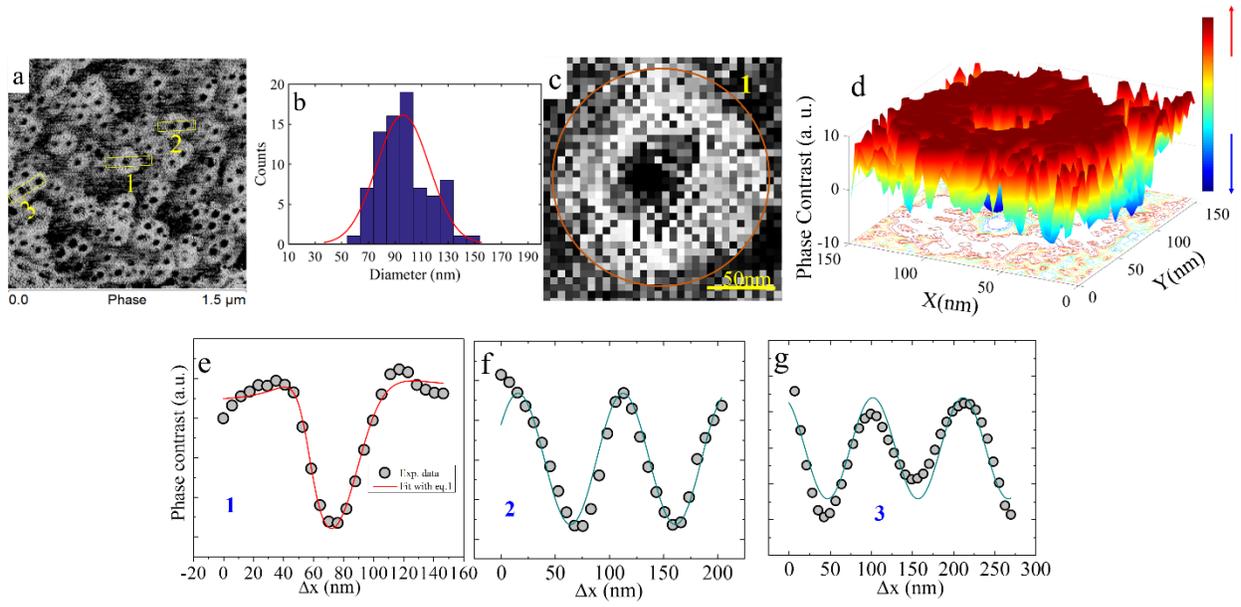

**Figure 2**

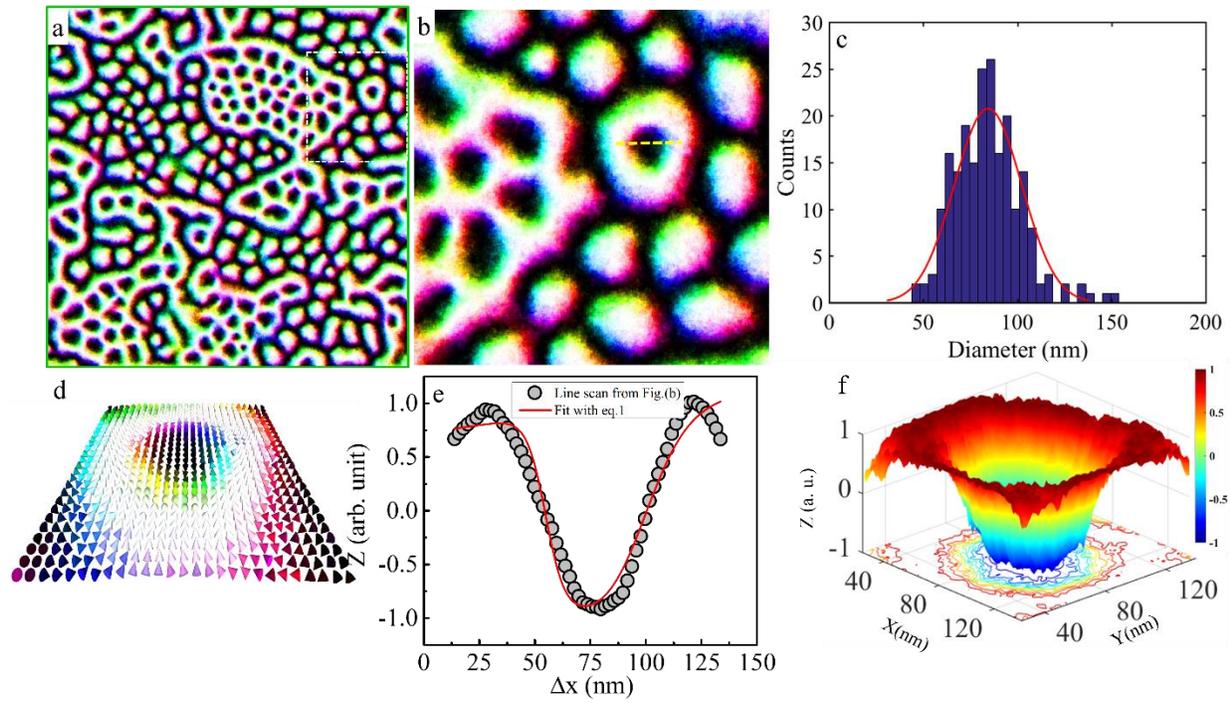

**Figure 3**

# Supplementary Information

## S1. Sample Preparation and Measurements

The trilayer thin film stacks consisting of Si(100)substrate/Ta(10)/CFA(10,5,1.8,1.0)/MgO (2) (numbers in parenthesis represent film thickness in *nm*) were *in-situ* deposited in sputter-up configuration using ion-assisted ion beam sputtering technique (after removing the native surface oxide layer of the Si substrates with HF (10:1 ratio) solution for 60s). Prior to deposition, the vacuum chamber was evacuated by a turbo molecular and cryo-pumps to a base pressure lower than ~$1\times10^{-7}$ Torr. A buffer layer of Ta(10 nm) was grown at room temperature and *in situ* annealed at 773K for 30 min in order to achieve a very flat surface. CFA layers with different thicknesses were then grown at room temperature followed by a thin layer of MgO on top of the CFA layer. During the growth of the Ta buffer and CFA layers, a working pressure of ~$8.5\times10^{-5}$ Torr was maintained during sputtering by flowing 4 sccm of Ar gas. This Ar was fed through a high energy RF-ion source comprising of two-grid assembly for extraction of ~4.5" dia Ar-ion beam. The high energy RF ion source was operated at 75 W with inner grid voltage $V_+$ = 500 V (i.e., Ar ion-energy) and the outer grid voltage V- = -270 V (for beam extraction/acceleration). The high-energy Ar ion beam so extracted was incident at an angle of 45° on 6" dia target mounted on a water-cooled target turret. A maximum of four targets, each of 6" dia can be mounted on the 4 sides of the turret which is rotatable *in-situ* to position any of the targets under the Ar-beam for sputtering. The target-substrate distance was ~27 cm.

For growing the MgO film, while the Mg target was sputtered using high energy RF-ion source (RF power=75 W, $V_+$ = 500 V and V- = -270V, Ar flow rate=3 sccm), the growing film was simultaneously irradiated (or to say *assisted*) with a low energy (50 eV) oxygen ion beam (at 45° on the growing film). This oxygen beam was extracted from a low energy RF-ion source (operated at RF power=75 W, $V_+$ = 50 V and V- = -30V, oxygen flow rate=12 sccm). Accordingly during the growth of MgO, whereas the total working pressure was slightly higher ~$1.0\times10^{-4}$ Torr, the $O_2$ partial pressure was ~$1.5\times10^{-5}$ Torr.

The deposition rates during the growth of Ta, CFA and MgO were 0.03nm/sec, 0.03nm/sec and 0.02nm/sec, respectively. The deposition rates were accurately calibrated individually and then also confirmed after deposition of the stack using data from X-Ray reflectivity measurements. Subsequent to the deposition of all the 3 layers, the trilayer stack was *in situ* post-annealed at a temperature of 523K for 1 hour in high vacuum ($6\times10^{-7}$ Torr).

Imaging of skyrmions in these trilayer thin films was performed in the remnant state (zero external magnetic field) by employing the *Bruker* make MFM (Model - Dimension *icon* with *ScanAsyst*). Magnetization vs. magnetic field (M-H) was recorded using a Quantum Design SQUID magnetometer in both *in-plane* and *out-of-plane* configurations. The thicknesses of individual layers, their density, the interface width and surface roughness were accurately estimated by simulating the X-ray reflectivity (XRR) profiles. The chemical valence state and the interface hybridization were examined by employing X-ray photoelectron spectroscopy (XPS). The XPS spectra on Ta(10)/CFA(1.8)/MgO(2) trilayer system were recorded using a *SPECS* make system with an Al-$K_\alpha$ x-ray source (1486.6 eV) and a hemispherical energy analyzer with pass energy of 40 eV and a resolution of ~0.7eV. Brillouin Light Scattering (BLS) spectra were recorded using a Sandercock-type (3+3) pass tandem Fabry-Pérot interferometer and a *p*-polarized (wavelength of 532nm with 300mW power) single longitudinal mode solid state laser. The details of the BLS set up can be found elsewhere[1]. The spectra were recorded in the conventional backscattering geometry at various wave vector orientations selected by mounting the sample on the angle controlled sample holder providing a range of 10°-60° incident angles corresponding to wave vectors $k_x$ lying in the range of 0.004-0.0204 nm$^{-1}$. The free spectral range of 50 GHz and a 29 multi-channel analyzer have been used for recording the spectra. An in-plane magnetic field of 0.1T was applied during measurements. All measurements were performed at room temperature.

**S2: *Experimental* Skyrmions in Ta(10)/CFA (1.0)/MgO (2) ultrathin films: MFM imaging**

The MFM, a non-contact scanning probe, is indispensable in studies of morphology and structure in terms of magnetic domains or magnetic nanostructures[2,3]. It is an important analytical tool whenever the near-surface stray-field variation of a magnetic sample is of interest. The MFM imaging was performed on the present samples in order to determine their remnant states after saturating the magnetization with a perpendicular field of 0.5T. A Co-Cr coated $Si_3Ni$ cantilever tip was used to map out the magnetic spin structure with a lateral resolution of 30 nm. The MFM imaging performed in this work favors the study of perpendicularly oriented magnetic domain structures. The MFM measurement consists of two steps, first tapping mode (topographical information) followed by lift mode (for magnetic domain information; positioned at a height of 70 nm from the film surface). The tip of the magnetic probe was magnetized parallel to the downward vertical direction using a permanent magnet. The resonant frequency of the cantilever (spring constant = 2.8 Nm$^{-1}$) was 73 kHz.

Fig. S1 shows MFM images of the sample Ta(10)/CFA(1.0)/MgO(2) recorded in remnant state. The images reveal well-developed skyrmions at room temperature. The signature of skyrmions is not clearly visible in Fig. S1(a) due to the large scan size. Zoomed (scanning) in portions of the image are shown in Figs. (b-d) revealing the formation of skyrmions together with a few strip-like domains which indicate that the system exhibits perpendicular magnetic anisotropy (PMA). The well-developed skyrmionic feature is clearly visible in Fig. S1(d) (the dotted line highlights the typical size of a skyrmion).

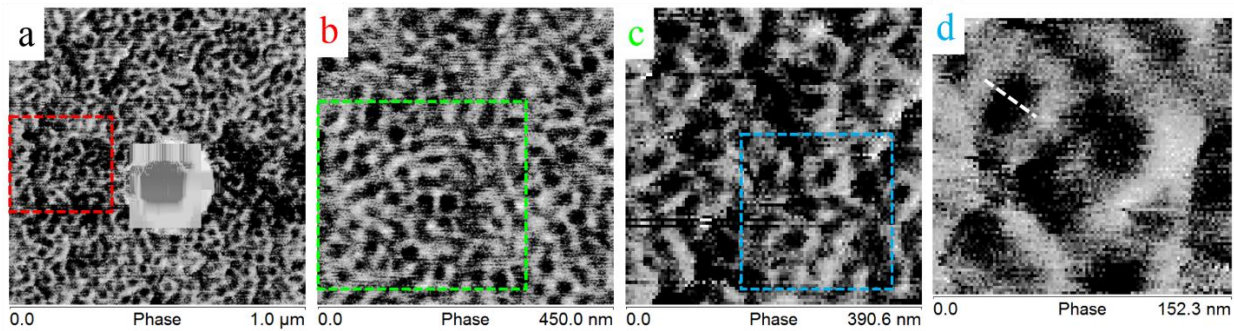

Figure S1: (a) MFM image revealing the formation of skyrmions in a Ta(10 nm)/CFA(1.0 nm)/ MgO(2.0 nm) thin film heterostructure at room temperature in the remnant state achieved from +0.5 T out-of-plane field. The large bright spot observed in (a) is due to the presence of a dust particle on the film surface. Using line scan profiling of a single skyrmion (dotted line in Fig. d)) the size/diameter of the skyrmion is estimated to be ~50nm by fitting using equation (2) (c.f. main text).

Further, the skyrmions of opposite polarity were also visualized by reaching the remnant state from a large negative field (-0.5T). Fig. S2 shows the MFM images of the Ta(10)/CFA(1.0)/MgO(2) thin film after magnetic saturation in a negative field. It is clearly seen that the polarity of skyrmions is reversed. Moreover, the line scan profile is shown in Fig. S2(d) indicates opposite contrast in the remnant state reached after saturation in a negative field. Here, it is also realized that the domain boundaries are not distinctly resolved because of the limited MFM lateral resolution, but the skyrmionic-state is identical even in the case of a multi-skyrmions, which signifies that the structure is topologically invariant. Thus, the skyrmion formation in Ta/CFA/MgO thin films is robust against defects and shape imperfections.

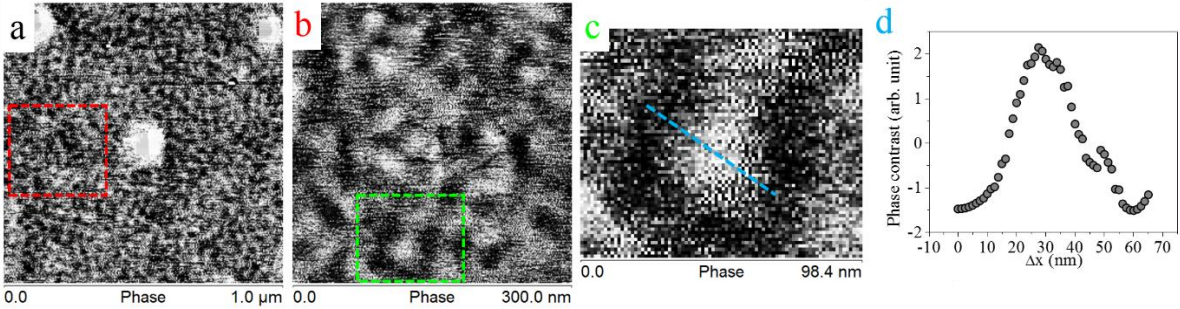

Figure S2: (a) Magnetic skyrmions in a Ta(10 nm)/CFA(1.0 nm)/MgO(2.0 nm) thin film heterostructure at room temperature in the remanent state but achieved from -0.5 T out-of-plane field, i.e., from a negatively saturated state. White spots in (a) are due to dust particles on the film surface. (b) and (c) are zoomed (scan) in areas of parts marked colored squares in (a) and (b), respectively. (d) The phase-contrast variation over a single skyrmion, i.e., across the yellow line scan in (c). The line scan profile on the single skyrmion (yellow line in (c)) indicates that the skyrmion has been nucleated with opposite polarity as compared to Fig. S1 where the sample was initially magnetized to saturation in a positive field.

**Micromagnetic Simulations:**

**Skyrmions in ultrathin films**

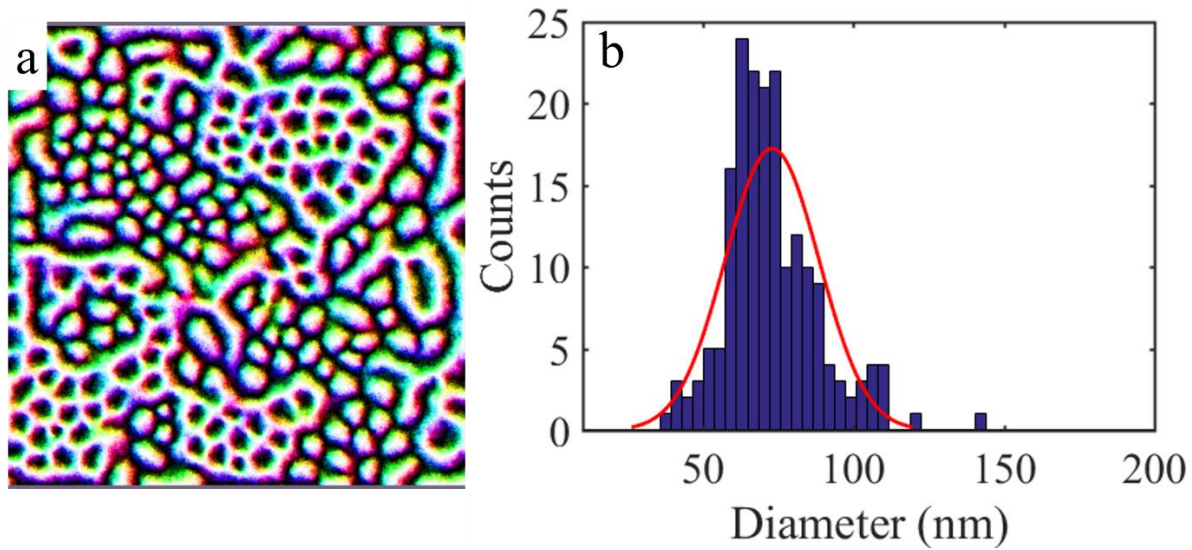

Figure S3: Micromagnetically simulated stable skyrmions in Ta(10)/CFA(1)/MgO(2) thin film stack of area 2 × 2 µm², keeping $K_u$ and $i$-DMI same as in the other simulations. (b) Statistical distribution of the skyrmions diameter.

**Comparison of skyrmion diameter (experimental and simulated)**

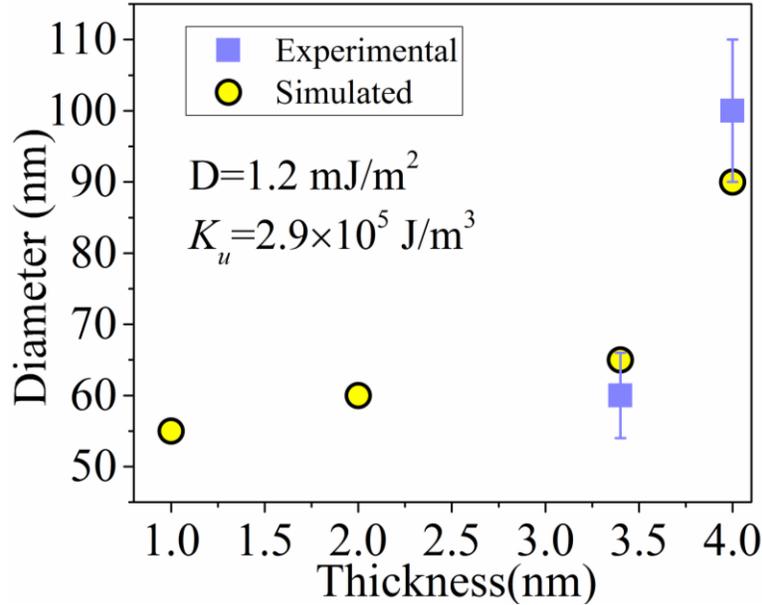

Fig. S4. Experimental and simulated skyrmion diameters in Ta/CFA/MgO thin films vs. effective FM layer thickness. Error bars in the experimental data is due to the standard deviation in at least 10 individual measurement using line scanning profile. The simulation was also done for lower thicknesses i.e., 1.0 and 2.0 nm and diameter plotted herewith (simulated images not shown here). The variation of the diameter is indicating that the tuning of the FM layer thickness can be useful for the fabrication of the skyrmion within the sub-50 nm diameter.

**S3: Interfacial Hybridization at CFA/MgO interface (Origin of SOC at interface for PMA)**

It is well known that the perpendicular magnetic anisotropy (PMA) in FM transition metals (TM) capped with MgO is attributed to the hybridization of O 2$p$ orbitals with either Co 3$d$ or Fe 3$d$ orbitals at the TM/MgO interface[4,5]. Here, we present the evidence of PMA in CFA/MgO whose origin is attributed to the interfacial interaction between the FM and the adjacent oxide layers. The interaction between these two layers provides the hybridization which leads to enhanced spin-orbit coupling (SOC) at the CFA/MgO interface. To ascertain such hybridization in the present case, XPS spectra were recorded on the CFA(1.8)/MgO(2) thin film stack. The XPS spectra were corrected using the C 1$s$ peak and deconvoluted using the XPS-peak 4.1 software. Fig. S5(a) shows the Mg 2$p$ spectrum where the peak at 50.6 eV designates the Mg$^{2+}$ state corresponding to MgO with no signal corresponding to Mg-metal. This ensured the formation of stoichiometric MgO during the deposition of the MgO layer. In Fig. S5(b) the asymmetric O 1$s$ peak is deconvoluted into two component peaks centered at 531.7 eV(OI) and 529.6 eV(OII), respectively coming from the interstitial O$^-$ state corresponding to the formation of MgO and CoO at the interface[6,7,8]. The Co 2$p$ electronic spectrum shown in Fig. S5(c) is also shown deconvoluted into two peaks, namely the Co 2$p$ as metal and Co 2$p$ as CoO at 780.04 eV and 778.35 eV, respectively. Thus, the presence of interfacial hybridization (which is essential for large SOC) of O 2*p* orbitals with Co 3*d* orbitals at the CFA/MgO interface is clearly evidenced in these samples.

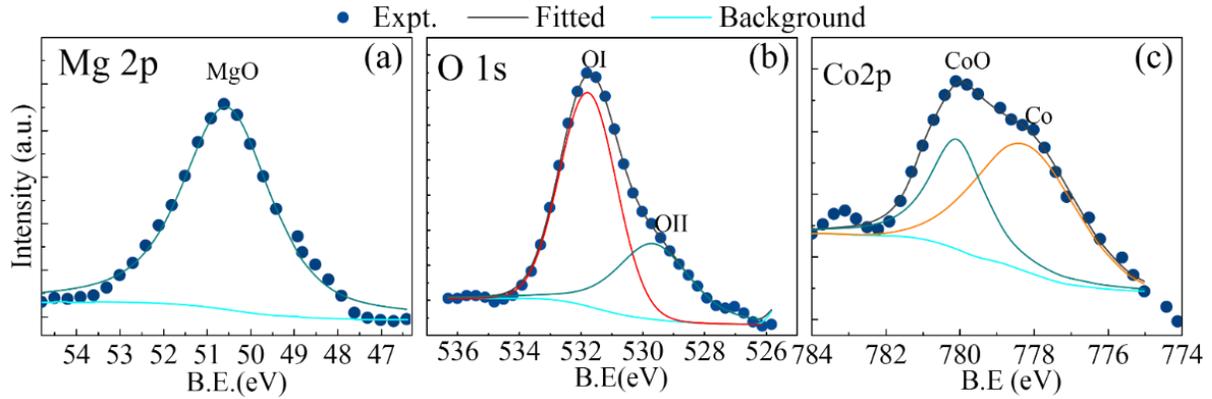

Figure S5: XPS spectra recorded on the Ta(10)/CFA(1.8)/MgO(2) thin film: (a) The MgO 2p peak shows the complete oxidation of Mg because of the absence of the Mg metal peak. (b) The interstitial O$^-$ ions peak appear due to interstitial oxygen in MgO (OI) and the shoulder indicates the metal-oxide formation (OII). (c) The clear observation of the CoO 2p$_{3/2}$ peak establishes the presence of the interfacial hybridization at the CFA-MgO interface.

## S4. X-ray reflectivity for thickness and interface width measurement

Figure S6 shows the simulated (solid line) and the experimental (data symbol) specular XRR spectra of the CFA thin films of two thicknesses. The simulated parameters are shown in Table.1. The XRR spectra illustrate the distinct presence of *Kiessig* fringes over the entire range of incident angle clearly indicating the presence of sharp and high quality interfaces.

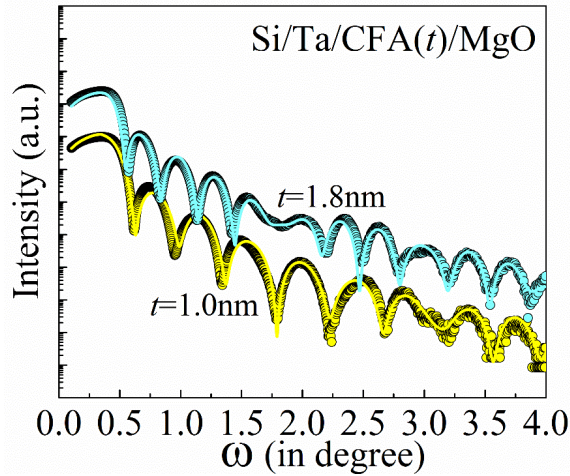

Figure S6: Specular X-ray reflectivity spectra of Si/Ta(10)/CFA(*t*)/MgO(2) thin films. The symbols are experimental data and the solid lines represent the simulated spectra obtained by a segmented fitting model. The simulated parameters are tabulated in Table.1.

**Table. 1:** Simulated parameters of Si/Ta(10)/CFA(*t*)/MgO(2) thin film samples.

| | *t*=1.8nm | | | *t*=1.0nm | | |
|---|---|---|---|---|---|---|
| **Layer** | $\rho$(gm/cc)±0.06 | *t*(nm)±0.03 | $\sigma$(nm) ±0.03 | $\rho$(gm/cc)±0.06 | *t*(nm)±0.02 | $\sigma$(nm) ±0.03 |
| SiO$_2$ | 02.00 | 02.55 | 0.21 | 02.30 | 02.47 | 0.34 |
| Ta | 12.67 | 10.95 | 0.50 | 13.44 | 10.11 | 0.37 |
| CFA | 07.50 | 01.68 | 0.28 | 05.05 | 01.00 | 0.36 |
| MgO | 02.50 | 02.28 | 0.48 | 02.00 | 02.50 | 0.25 |
| MgO$_x$ | 02.00 | 01.42 | 0.33 | 01.01 | 01.36 | 0.23 |

### S5: Magnetization measurement

Figure S7 shows the magnetization (*M*) vs. applied magnetic field (H) for the Ta(10)/CFA(1.8)/MgO(2) trilayer film recorded both in the *in-plane* and the *out-of-plane* magnetic field orientations. The latter M-H loop exhibits lower saturation field compared to the former suggesting that the CFA film possesses PMA at room temperature. The saturation magnetization is found to be 838 kA/m which is close to the value of 1000 kA/m reported for the bulk CFA Heusler alloy[9]. The perpendicular uniaxial magnetic anisotropy energy, evaluated from the difference in the measured values of saturation fields, was found to be ~2.9 ×10$^5$ J/m$^3$. From the magnetization curves, it is observed that even though the film possesses PMA at room temperature the hysteresis and the remnant magnetization $M_r$ both were negligibly small which are characteristic features of the formation of the chirality in the remnant state as explained by Cowburn *at al*[10]. Thus, the magnetization measurements at room temperature provide first-hand experimental evidence of the formation of a topological state possessing chirality, possibly as a clue to the presence of skyrmions at room temperature in the remnant state.

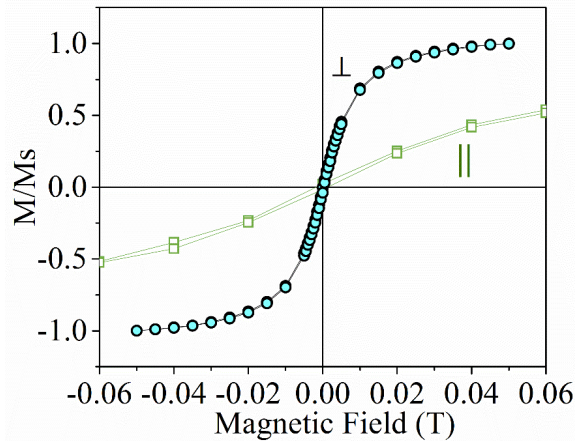

FigureS7: (color online) Magnetization vs. magnetic field loops recorded for the Ta(10)/ CFA(1.8)/MgO(2) sample at room temperature for two magnetic field orientations, in-plane (∥) and out-of-plane (⊥).

## S6: Micromagnetic Simulation

The micromagnetic simulations were performed using the mumax 3.8 simulation tool[11] integrated with the Landau-Lifshitz-Gilbert (LLG)

$$\frac{\partial M_s}{\partial t} = -\gamma M_s \times H_{eff} + \alpha M_s \times \frac{\partial M_s}{\partial t},$$

where γ is the gyromagnetic ratio, $\alpha$ is the Gilbert damping constant, $M_s$ is the saturation magnetization and $H_{eff}$ is the local effective field consisting of as the following contributions

$$\begin{aligned}H_{eff} &= H_{exc} + H_{ani} + H_{dmg} + H_{ext} + H_{dmi}\\ &= \frac{2A_{ex}}{\mu_0 M_s}\left[(\partial m_x/\partial x)^2 + (\partial m_y/\partial y)^2 + (\partial m_z/\partial z)^2\right] + \frac{2K_u}{\mu_0 M_s} - \mu_0^{-1} M_s(m_x N_x + m_y N_y + m_z N_z)\\ &\quad + H_{ext} + \frac{2D}{\mu_0 M_s}\left(\frac{\partial m_z}{\partial x} + \frac{\partial m_z}{\partial y} - \left(\frac{\partial m_x}{\partial x} + \frac{\partial m_y}{\partial x}\right)\right)\end{aligned}$$

$H_{iDM}$ is the field due to DM interaction which is primarily responsible for curling of the magnetization and for skyrmion formation. In addition to above terms, a thermal fluctuating field $H_{TH}$ is included which controls the temperature effect in the micromagnetic system. It is modeled by a Langevin random field $H_L(H_{L_x}, H_{L_y}, H_{L_z})$ where each component follows a zero-mean Gaussian random process whose standard deviation is a function of temperature which is defined by $\delta = \sqrt{2\alpha k_B T / \gamma \mu_0 M_s v_F \Delta t}$ ($k_B$ is a Boltzmann constant, $v_F$ is the micromagnetic cell volume, $T$ is the temperature, and $\Delta t$ is the integration time-step.

For simulation of the extended thin film, a square shaped geometry with dimensions 2048 nm×2048 nm was considered. The finite element computational cell/grid size was kept constant at 2 nm as rectangular discrete grids with constant thickness of 1−4 nm considered to be effective thickness, i.e., the sum of the thicknesses of FM and the adjoining layer and is assumed to be optimum for the relevant micromagnetic length scales[12]. The damping constant was kept fixed to 0.01. The skyrmionic state was observed in particular combinations of parameters which were determined with the help of our experimental results.

The simulations were performed starting with an initial random magnetization and then relaxed the system (both in zero magnetic field), which eventually resulted in a magnetization configuration corresponding to the skyrmions state. However, when the relaxation to eventual magnetization configuration was simulated using an initial state corresponding to different out-of-plane magnetic field (i.e., non-remnant state), it was observed that the skyrmions start disappearing on above 100mT. The simulations show that for fields above 500mT, the final state corresponds to ferromagnetic continuum state.

**References:**

1. Chaurasiya, A. K. *et al.* Direct observation of interfacial Dzyaloshinskii-Moriya interaction from asymmetric spin-wave propagation in W/CoFeB/SiO$_2$ heterostructures down to sub-nanometer CoFeB thickness. *Sci. Rep.* **6,** 32592 (2016).

2. Zhu, X. & Grütter, P. I maging , manipulation , and spectroscopic measurements of nanomagnets by magnetic force microscopy. *MRS Bull.* 457–462 (2004).

3. Shinjo, T. magnetic vortex core observation in circular dots of permalloy. *Science* **289,** 930–932 (2000).

4. Yang, H. X. *et al.* First-principles investigation of the very large perpendicular magnetic anisotropy at Fe|MgO and Co|MgO interfaces. *Phys. Rev. B* **84,** 54401 (2011).

5. Baumann, S. *et al.* Origin of perpendicular magnetic anisotropy and large orbital moment in Fe atoms on MgO. *Phys. Rev. Lett.* **115,** 237202 (2015).

6. Singh, B. B., Agrawal, V., Joshi, A. G. & Chaudhary, S. X-ray photoelectron spectroscopy and conducting atomic force microscopy investigations on dual ion beam sputtered MgO ultrathin films. *Thin Solid Films* **520,** 6734–6739 (2012).

7. Zhang, J. Y. *et al.* Effect of MgO/Co interface and Co/MgO interface on the spin dependent transport in perpendicular Co/Pt multilayers. *J. Appl. Phys.* **116,** 163905 (2014).

8. Lopez-Santiago, A. *et al.* Cobalt ferrite nanoparticles polymer composites based all-optical magnetometer. *Opt. Mater. Express* **2,** 978 (2012).

9. Husain, S., Akansel, S., Kumar, A., Svedlindh, P. & Chaudhary, S. Growth of Co$_2$FeAl Heusler alloy thin films on Si(100) having very small Gilbert damping by Ion beam sputtering. *Sci. Rep.* **6,** 28692 (2016).

10. Cowburn, R. P., Koltsov, D. K., Adeyeye, a. O., Welland, M. E. & Tricker, D. M. Single-domain circular nanomagnets. *Phys. Rev. Lett.* **83,** 1042 (1999).

11. Vansteenkiste, A. *et al.* The design and verification of MuMax3. *AIP Adv.* **4,** 107133 (2014).

12. Woo, S. *et al.* Observation of room-temperature magnetic skyrmions and their current-driven dynamics in ultrathin metallic ferromagnets. *Nat. Mater.* **15,** 501–506 (2016).